\documentclass[12pt]{iopart}
\usepackage{iopams}
\usepackage{graphicx}
\begin{document}

\title[Magnetization and susceptibility of ferrofluids]
{Magnetization and susceptibility of ferrofluids}

\author{I Szalai$^1$, S Dietrich$^{2,3}$}

\address{$^1$ Institute of Physics, University of Pannonia, H-8201 Veszpr\'em, PO Box 158, Hungary}
\address{$^2$ Max-Planck-Institut f\"ur Metallforschung, Heisenbergstr. 3, D-70569 Stuttgart, Germany}
\address{$^3$ Institut f\"ur Theoretische und Angewandte Physik, Universit\"at Stuttgart, 
Pfaffenwaldring 57, D-70569 Stuttgart, Germany}

\ead{szalai@almos.vein.hu}

\begin{abstract}
A second-order Taylor series expansion of the free energy functional provides 
analytical expressions for the magnetic field dependence 
of the free energy and of the magnetization of ferrofluids, here modelled by dipolar 
Yukawa interaction potentials. The corresponding hard core dipolar Yukawa reference 
fluid is studied within the framework of the mean spherical approximation. 
Our findings for the 
magnetic and phase equilibrium properties are in quantitative agreement with 
previously published and new Monte Carlo simulation data.
\end{abstract}

\section{Introduction}
Ferrofluids are colloidal suspensions of ferromagnetic grains dispersed in a solvent. 
To keep such dispersions stable, one must avoid aggregation and counterbalance the 
attractive van der Waals interaction by repulsive forces, either electrostatic ones, 
if the particles are charged, or steric ones, if their surface is 
coated with polymers or surfactants. 
Each particle of a ferrofluid possesses a permanent magnetic dipole moment so that 
the ferrocolloids have much in common with dipolar liquids \cite{bu,hu}. 
The effective interaction of such magnetic 
particles can be modeled by the Stockmayer potential, which is the sum of an isotropic 
Lennard-Jones (LJ) and the anisotropic dipole-dipole interaction \cite{ba,g1}. 
Within this approach the LJ interaction parameters are considered to be effective ones, 
incorporating the effects of the solvent. (In a more refined approach one would model the 
suspension as a binary mixture of a nonpolar solvent and a polar solute component 
\cite{sa1,kl1}.)

Based on previous experience \cite{h1,s1}, here we substitute the 
Lennard-Jones potential by a hard core Yukawa potential. 
For a homogeneous isotropic dipolar Yukawa (DY) fluid this substitution
provides the valuable benefit that one can solve analytically the corresponding 
Ornstein-Zernike equation within the mean spherical approximation (MSA). 
It turns out that the thermodynamic 
properties of a DY fluid can be calculated easily from the corresponding results for a 
hard core Yukawa and a dipolar hard sphere (DHS) fluid. The thermodynamic properties of 
a DY fluid differ from those of the DHS fluid but the magnetic susceptibility of the 
DY fluid is the same as that of the DHS fluid, which was predicted first by 
Wertheim \cite{w1}. The disadvantage of the MSA is that it can predict the magnetization 
only for weak magnetic fields because it is a linear response theory. 
In order to eliminate this shortcoming, first Lebedev \cite{l1} suggested a 
semi-empirical method for using MSA at finite magnetic fields.  
Later Morozov and Lebedev \cite{m1} extended the applicability of DHS MSA to the case 
of arbitrary strengths of the magnetic fields on the basis of the Lovett-Mou-Buff 
equation \cite{l2}, relating the single particle distribution function and the 
direct correlation function. Within the 
framework of density functional theory, here we also propose an equation for the 
magnetization of ferrofluids, which turns out to be simpler than the corresponding 
equation of Morozov and Lebedev \cite{m1}. In order to facilitate the calculation of 
the phase diagrams, we present also the magnetic field dependence of the chemical 
potential and of the pressure.       

\section{The model}

We study so-called hard core dipolar Yukawa fluids which consist of
spherically symmetric particles interacting via a Yukawa (Y) potential
characterized by parameters $\sigma$, $\epsilon$, and $\lambda$:
\begin{equation}
u_Y(r_{12})=-\frac{\epsilon\sigma}{r_{12}}\exp[-\lambda(r_{12}-\sigma)].
\label{Yu}
\end{equation}  
In addition there is a dipolar interaction due to point dipoles embedded at
the centres of the particles:
\begin{equation}
u_{D}(\mathbf{r}_{12},\omega_1,\omega_2)=-\frac{m^2}{r_{12}^3}
D(\omega_{12},\omega_1,\omega_2)),
\end{equation}
with the rotationally invariant function
\begin{equation}
D(\omega_{12},\omega_1,\omega_2)=
3[\widehat{\mathbf{m}}_1(\omega_1)\cdot\widehat{\mathbf{r}}_{12}]
[\widehat{\mathbf{m}}_2(\omega_2)\cdot\widehat{\mathbf{r}}_{12}]
-[\widehat{\mathbf{m}}_1(\omega_1)\cdot\widehat{\mathbf{m}}_2(\omega_2)],
\label{D}
\end{equation}
where particle 1 (2) is located at $\mathbf{r}_1$ ($\mathbf{r}_2$) and carries 
a dipole moment of strength $m$ with an orientation given by the unit vector 
$\widehat{\mathbf{m}}_1(\omega_1)$ ($\widehat{\mathbf{m}}_2(\omega_2)$) with polar angles 
$\omega_1=(\theta_1,\phi_1)$ ($\omega_2=(\theta_2,\phi_2)$); 
$\mathbf{r}_{12}=\mathbf{r}_1-\mathbf{r}_2$ is the difference vector between the
centres of particle 1 and 2, $r_{12}=|\mathbf{r}_{12}|$, and 
$\widehat{\mathbf{r}}_{12}=\mathbf{r}_{12}/r_{12}$ is a unit vector with 
orientation $\omega_{12}=(\theta_{12},\phi_{12})$.
The hard core dipolar Yukawa potential is defined by the aforementioned
interaction potentials as 
\begin{equation}
u_{DY}(\mathbf{r}_{12},\omega_1,\omega_2)=\left\{
        \begin{array}{lll}
        \infty &, & r < \sigma \\
      u_Y(r_{12})+u_{D}(\mathbf{r}_{12},\omega_1,\omega_2) &, & r \geq \sigma.
        \end{array} 
        \right. \
\label{uDY}
\end{equation}
In the following we consider the DY fluid in a homogeneous external magnetic 
field $H$ (the direction of which is taken to coincide with the direction of the 
$z$ axis).
The contribution of the latter to the interaction potential is
\begin{equation}
u_{ext}(\omega)=-mH\cos\theta,
\end{equation}
where the angle $\theta$ measures the orientation of the dipole of a single particle 
relative to the field direction.

\section{Homogeneous magnetization in an external field}

In general, an inhomogeneous and anisotropic dipolar fluid is described 
by the one particle distribution function $\hat\rho(\mathbf{r},\omega)$. 
For bulk fluid phases 
the number density $\rho=\int{d\omega}\hat\rho(\mathbf{r},\omega)$ is
spatially constant and  
$\alpha(\mathbf{r},\omega)=\hat\rho(\mathbf{r},\omega)/\rho$ is the 
orientational distribution function at the point $\mathbf{r}$. 
In a homogeneously magnetized bulk phase the orientational distribution function 
depends only on $\omega$ ($\alpha(\mathbf{r},\omega)=\alpha(\omega)$) and it is 
normalized to 1, i.e., $\int{d\omega}\alpha(\omega)=1$. 
In these terms our subsequent bulk analysis is based on the minimization of the 
following grand canonical variational functional: 
\begin{equation}
\Omega[\rho,\{\alpha(\omega)\},T,\mu]=
F_{DY}[\rho,\{\alpha(\omega)\},T]
-\int{d^3rd\omega\rho\alpha(\omega)(\mu-u_{ext}(\omega))},
\end{equation}
where $F_{DY}$ is the Helmholtz free energy functional of the DY fluid, $T$ 
is the temperature and $\mu$ is the chemical potential. The Helmholtz free energy 
functional consists of an ideal gas and an excess contribution:
\begin{equation}
F_{DY}[\rho,\{\alpha(\omega)\},T]=F^{id}[\rho,\{\alpha(\omega)\},T]
+F_{DY}^{ex}[\rho,\{\alpha(\omega)\},T].
\end{equation}
For the system under study the ideal contribution has the form
\begin{equation}
F^{id}[\rho,\{\alpha(\omega)\},T]=
V\beta^{-1}\rho\Big[\ln(\rho\Lambda^3)-1
+\int{d\omega\alpha(\omega)\ln(4\pi\alpha(\omega))}\Big],
\end{equation}
where $\Lambda$ is the de Broglie wavelength, $\beta=1/(k_BT)$ is the
inverse temperature and $V$ is the volume of the fluid. 
The central quantity in this
theory is the excess free energy functional $F^{ex}$, which originates from
interparticle interactions. The excess DY free energy functional for an 
{\it a}nisotropic system is not known. 
However, it can be approximated by a  functional Taylor series, expanded around
a homogeneous {\it i}sotropic reference system with bulk density $\rho$. 
Neglecting all terms beyond second order one has   
\begin{eqnarray}
\beta{F^{ex,a}_{DY}[\rho,\{\alpha(\omega)\},T]}=
\beta{F^{ex,i}_{DY}(\rho,T)}
-\rho\int{d^3r_1d\omega\Delta\alpha(\omega)c^{(1)}_{DY}(\rho)}\nonumber\\
-\frac{\rho^2}{2}\int{d^3r_1d\omega_1}\int{d^3r_2d\omega_2}\Delta\alpha(\omega_1)
\Delta\alpha(\omega_2)c^{(2)}_{DY}(\mathbf{r}_{12},\omega_1,\omega_2,\rho),
\label{fex}
\end{eqnarray}
where $\Delta\alpha(\omega)=\alpha(\omega)-1/(4\pi)$ is the difference 
between the anisotropic and the isotropic orientational distribution function. 
The first- ($c^{(1)}$) and second-order ($c^{(2)}$) direct correlation 
functions appearing in Eq. (\ref{fex}) are the first 
and second functional derivatives of ($-\beta{F^{ex,a}_{DY}}$) evaluated at the 
homogeneous and isotropic density $\rho$. Since $\int{d\omega\Delta\alpha(\omega)=0}$
only the second- and higher-order direct correlation functions provide nonzero 
contributions to the functional $\beta{F^{ex,a}_{DY}}$. 
In the following we approximate the second-order direct
correlation function of a DY fluid by the corresponding MSA solution
due to Henderson {\it{et al}} \cite{h1}:
  
\begin{equation}
\!\!\!\!\!\!\!\!\!\!\!\!\!\!\!\!\!\!\!\!\!\!\!\!\!\!\!\!\!\!\!\!
c^{(2)}_{DY}(\mathbf{r}_{12},\omega_1,\omega_2,\rho)=
c^{(2)}_Y(r_{12},\rho)
+c^{(2)}_{D}(r_{12},\rho)D(\omega_{12},\omega_1,\omega_2)
+c^{(2)}_{\Delta}(r_{12},\rho)\Delta(\omega_1,\omega_2)
\label{cor}
\end{equation}
(see Eq. (\ref{D})) with the rotationally invariant function
\begin{equation}
 \Delta(\omega_1,\omega_2)=
\widehat{\mathbf{m}}_1(\omega_1)\cdot\widehat{\mathbf{m}}_2(\omega_2) .
\end{equation}
In Eq. (\ref{cor}), where $c^{(2)}_Y$ is the hard core Yukawa direct 
correlation function \cite{ha1}, 
$c^{(2)}_D$ and $c^{(2)}_{\Delta}$ are spherical harmonic components 
of the corresponding direct correlation function of a DHS fluid in MSA \cite{w1}. 
The latter two functions can be expressed in terms of the Percus-Yevic hard sphere 
direct correlation function. If the volume
of the fluid $V$ has the shape of a rotational ellipsoid (elongated around the magnetic
field direction) and the magnetization is homogeneous, $\alpha(\omega)$ depends only on 
the angle $\theta$ relative to the long axis of the ellipsoid, and thus it can be 
expanded in terms of Legendre polynomials: 
\begin{equation}
\alpha(\omega)=\frac{1}{2\pi}\overline\alpha(\cos\theta)=
\frac{1}{2\pi}\sum_{i=0}^{\infty}\alpha_iP_i(\cos\theta).
\end{equation} 
Due to $\alpha_0=1/2$ one has  
$\Delta\alpha(\omega)=\frac{1}{2\pi}\sum_{i=1}^{\infty}\alpha_iP_i(\cos\theta)$.
In order to avoid depolarization effects, we consider sample shapes of infinitely 
prolate ellipsoids, i.e., needle-shaped volumes $V$. Since in an expansion in terms of 
$P_i(\cos\theta)$ the orthogonal functions $D$ and $\Delta$ (see Eq. (\ref{cor})) 
have contributions 
only of the order $i\leq1$, $\alpha_1$ is the only term in 
$\overline\alpha(\cos\theta)$ which contributes to Eq. (\ref{fex}). 
Thus after some elementary calculations one finds for the free energy 
functional of Eq. (\ref{fex})    
\begin{equation}
\frac{F^{ex,a}_{DY}[\rho,\{\alpha(\omega)\},T]}{V}=
f^{ex}_{DY}(\rho,T)
-\frac{2\rho}{3\beta}(1-q(-\xi))\alpha_1^2,
\end{equation}
where $f^{ex}_{DY}=F^{ex,i}_{DY}/V$ is the Helmholtz free energy 
density of the isotropic DY fluid and 
\begin{equation}
q(x)=\frac{(1+2x)^2}{(1-x)^4}
\end{equation} 
is the reduced inverse compressibility function of hard spheres 
within the Percus-Yevic approximation \cite{pe1}. 
The parameter $\xi(\chi_L)$ stems from the DHS MSA and is given by the 
implicit equation
\begin{equation}
4\pi\chi_L=q(2\xi)-q(-\xi),
\label{sus}
\end{equation}
where $\chi_L=\beta\rho{m^2}/3$ is the Langevin susceptibility. Minimization
of the grand canonical functional with respect to the orientational distribution
function yields
\begin{equation}
\overline{\alpha}(\cos\theta)=C^{-1}
\exp\left[(\beta mH+2(1-q(-\xi))\alpha_1)P_1(\cos\theta)\right],
\end{equation}
with the normalization constant
\begin{equation}
C=\frac{2\sinh(\beta mH+2(1-q(-\xi))\alpha_1)}{\beta mH +2(1-q(-\xi))\alpha_1}.
\end{equation}
The coefficient $\alpha_1$ is determined by the implicit equation
\begin{equation}
\alpha_1=\frac{3}{2}L\left(\beta{mH}+2(1-q(-\xi))\alpha_1\right),
\label{al1}
\end{equation}
where $L(x)$ is the well known Langevin function. Further details of this minimization 
scheme can be found in Ref. \cite{g1}. Minimization of $\Omega/V$ with
respect to $\rho$ yields
\begin{equation}
\beta\mu=\beta\mu_{DY}(\rho,T)
-\ln[C/2]
+\frac{2}{3}\rho\alpha_1^2\frac{\partial q(-\xi)}{\partial\rho}
+\frac{2}{3}(1-q(-\xi))\alpha_1^2,
\label{chem}
\end{equation}
where $\mu_{DY}$ is the chemical potential of the DY fluid \cite{h1,s1}. 
By eliminating the chemical potential the equilibrium grand 
canonical potential can be expressed as 
\begin{equation}
\frac{\Omega(\rho,\{\alpha_1\},T)}{V}=f_{DY}(\rho,T)-\rho\mu_{DY}(\rho,T)
-\frac{2}{3}\frac{\rho^2}{\beta}\alpha_1^2\frac{\partial q(-\xi)}{\partial\rho},
\end{equation}
where $f_{DY}=F_{DY}/V$ is the free energy density of the homogeneous and 
isotropic DY fluid \cite{h1,s1}.
Due to $\Omega=-pV$ the corresponding pressure of an equilibrium phase 
as a function of the density and temperature is
\begin{equation}
p=p_{DY}(\rho,T)+\frac{2}{3}\frac{\rho^2}{\beta}\alpha_1^2\frac{\partial q(-\xi)}
{\partial\rho},
\label{pres}
\end{equation}
where $p_{DY}=-f_{DY}+\rho\mu_{DY}$ is the pressure of the DY fluid.
Each particle of the magnetic fluid carries a dipole moment which may be 
preferentially aligned in a particular direction. This gives rise to a 
magnetization $M$ (the direction of which coincides with the direction of the external 
magnetic field):
\begin{equation}
M=\rho\int{d\omega}\alpha(\omega)m\cos\theta=\frac{2}{3}m\rho\alpha_1.
\label{magn1}
\end{equation}
Equations (\ref{magn1}) and (\ref{al1}) lead to an implicit equation for
the external field dependence of the magnetization:
\begin{equation}
M=m\rho{L\left(\beta{mH}+3M\frac{(1-q(-\xi))}{m\rho}\right)}.
\label{magn2}
\end{equation}
Accordingly, the zero field magnetic susceptibility $\chi$ is
\begin{equation}
\chi=\frac{\chi_L}{q(-\xi)}.
\end{equation}
This result is identical with the ones of Henderson {\it et al} \cite{h1} 
for DY fluids and with Wertheim's \cite{w1} equation for DHS fluids.
\section{Monte Carlo simulations}

We carried out Monte Carlo simulations for DY fluids using NVT and NpT ensembles 
in order to verify the predictions of the present DFT. The initial 
configurations in a cubic box were created by random insertions of the particles, 
excluding particle overlaps. Boltzmann sampling and periodic boundary conditions 
with the minimum-image convention were applied. A spherical cutoff 
of the pair interaction potentials at half of the 
cell length was applied, and long-ranged corrections (LRC) were taken into account. 
For the dispersion part of the potential, the usual Yukawa-tail LRC was used while 
for the dipole-dipole interaction a reaction field LRC and a conducting boundary 
condition were applied. For the simulation of the magnetization after 40000 equilibration 
cycles 0.6-0.8 million production cycles were used involving 512 particles. 
In the simulations the equilibrium magnetization can be obtained from the
expression 
\begin{equation}
\mathbf{M}=\frac{1}{V}\left\langle\sum_{i=1}^N\mathbf{m}_i\right\rangle,
\end{equation}
where the brackets denotes the ensemble average.
 
We obtained the external field dependence of the numerical data for liquid-vapor 
coexistence of a DY fluid by using the extended NpT plus test particle (NpT+TP) 
method, which is described in detail in Ref. \cite{b2} so that here we present 
only an outline. 
Starting from a point $(T_0,p_0)$ in the one-phase region of the $(T,p)$ parameter 
plane, the chemical potential $\mu$ can be expanded into a Taylor series about 
the point $(T_0,p_0)$ up to third order:
\begin{equation}
\mu(T,p)=\mu(T_0,p_0)+\sum_{i=1}^3
\frac{1}{i!}\left[(T-T_0)\frac{\partial}{\partial{T}}+(p-p_0)
\frac{\partial}{\partial{p}}\right]^i{\mu(T,p)}.
\end{equation}

On the basis of basic thermodynamic relations the coefficients of this series can 
be expressed in terms of the derivatives of the enthalpy and the volume of the system 
with respect to $T$ and $p$ and they can be calculated from fluctuation formulas 
by performing an NpT+TP Monte Carlo simulation at the point $(T_0,p_0)$. 
These derivatives and fluctuation formulas are given in Ref. \cite{b2}. 
Carrying out this scheme for a point in the liquid and in the vapor phase, 
respectively, and rewriting the third-order Taylor series of 
$\mu$ for these points accordingly, the vapor pressure curve as well as other 
equilibrium data 
can be obtained from the intersection of these curves in the appropriate temperature range 
within an accuracy depending on the number of terms taken into account in the 
Taylor expansion. The NpT ensemble MC simulations involved 512 particles 
and about 0.8 million cycles were performed. The chemical potential 
was calculated by using Widom's test particle method.
\section{Results and discussion}
In the following we shall use reduced quantities: $T^*=kT/\epsilon$ as 
reduced temperature, $\rho^*=\rho\sigma^3$ as reduced density, 
$m^*=m/\sqrt{\epsilon\sigma^3}$ as reduced dipole moment, 
$H^*=H\sqrt{\sigma^3/\epsilon}$ as reduced magnetic field strength,
 and $M^*=M\sqrt{\sigma^3/\epsilon}$ as the reduced magnetization.
For the reference system, all results are obtained for $\lambda=1.8/\sigma$ 
(see Eq. (\ref{Yu})), and different from Ref. \cite{s1} 
the so-called first-order MSA \cite{t1} is used to describe the Yukawa fluid. 
The magnetization curves obtained from the numerical solution of Eqs. (\ref{magn2}) 
and (\ref{sus})
\begin{figure}[h]
\begin{minipage}{18pc}
\includegraphics*[width=17.8pc]{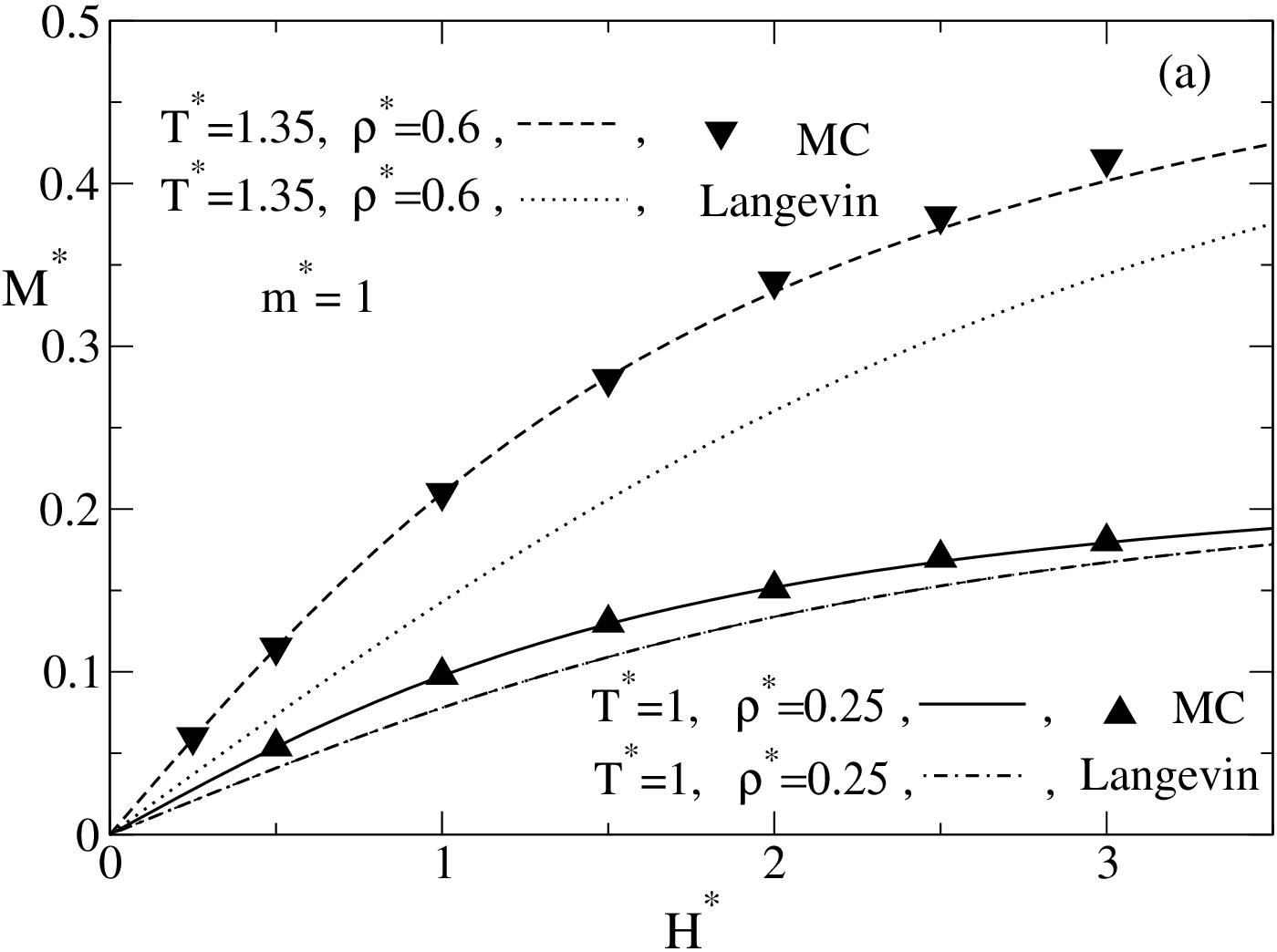}
\end{minipage}\hspace{0pc}%
\begin{minipage}{18pc}
\includegraphics*[width=18.6pc]{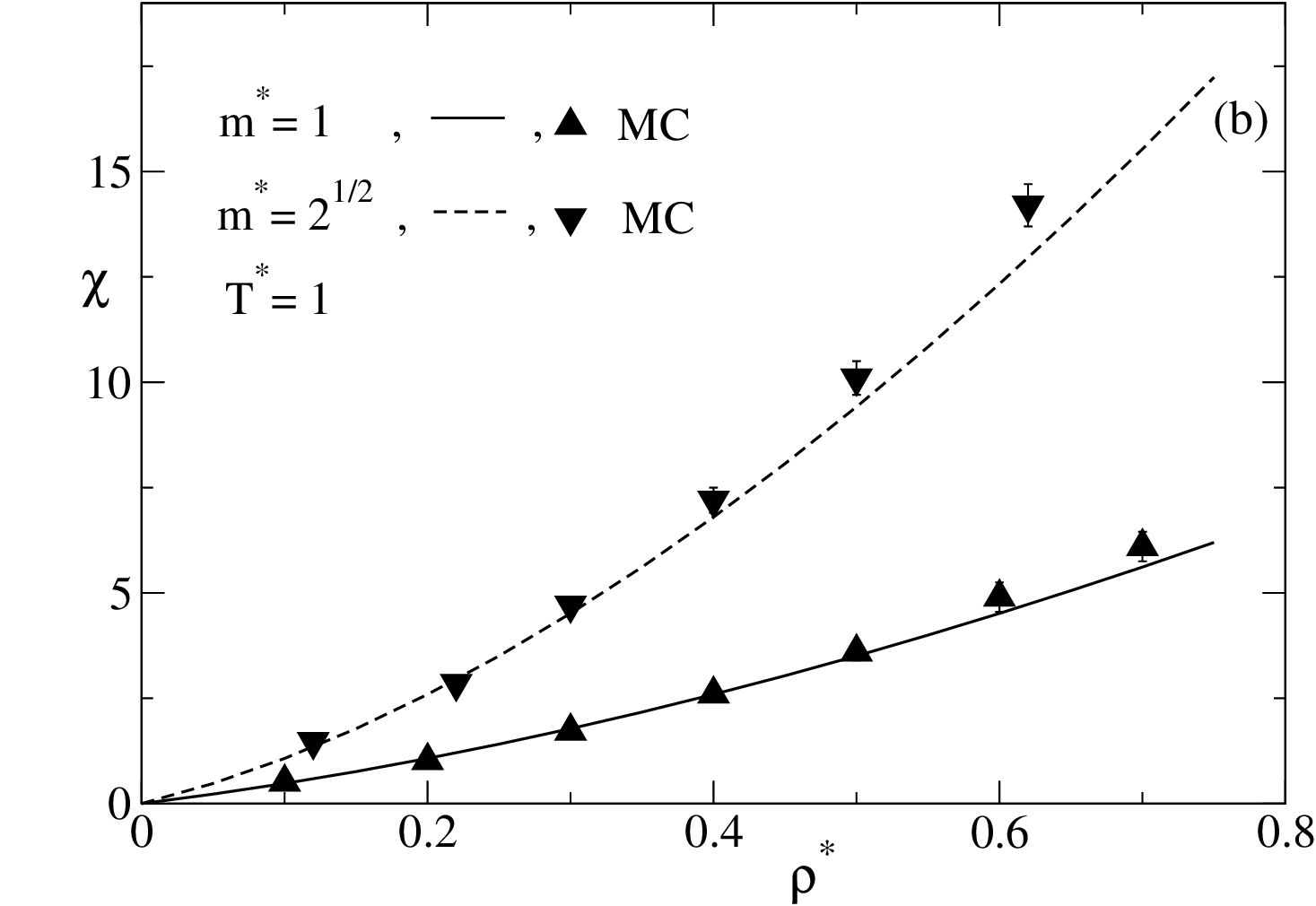}
\end{minipage}
\begin{minipage}{37pc}
\caption{\label{label}(a) Magnetization curves of DY fluids for different values 
of the reduced 
density $\rho^*$ and reduced temperature $T^*$ for $m^*=1$. 
(b) Magnetic zero-field susceptibility of 
DY fluids for different values of the reduced dipole moment and for the reduced 
temperature $T^*=1$. 
The symbols in both figures are the 
results of Monte Carlo simulations whereas the lines are obtained from the present DFT.} 
\end{minipage}
\end{figure}
\begin{figure}[h]
\begin{minipage}{18pc}
\includegraphics*[width=17.8pc]{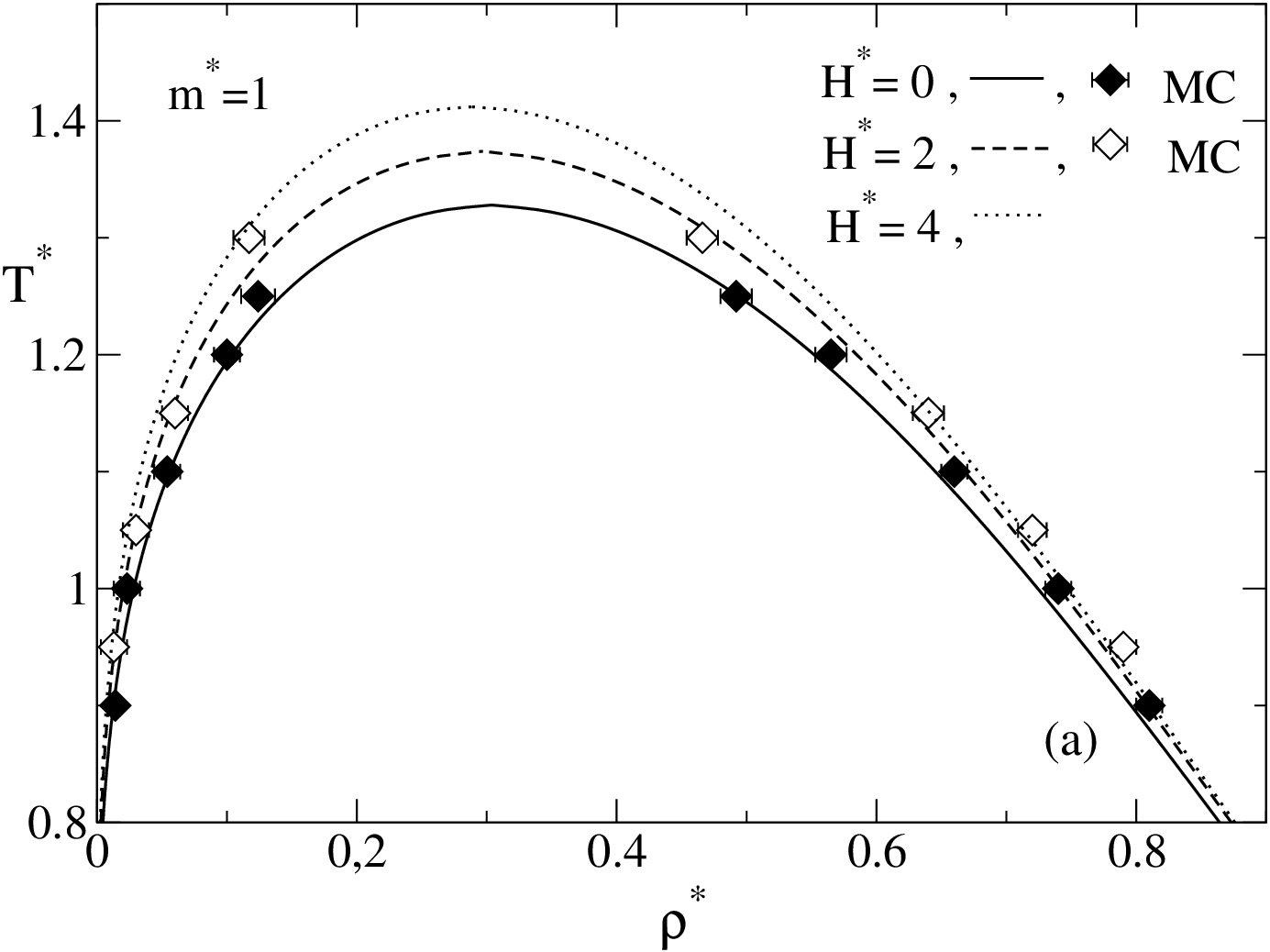}
\end{minipage}\hspace{0pc}%
\begin{minipage}{18pc}
\includegraphics*[width=18.6pc]{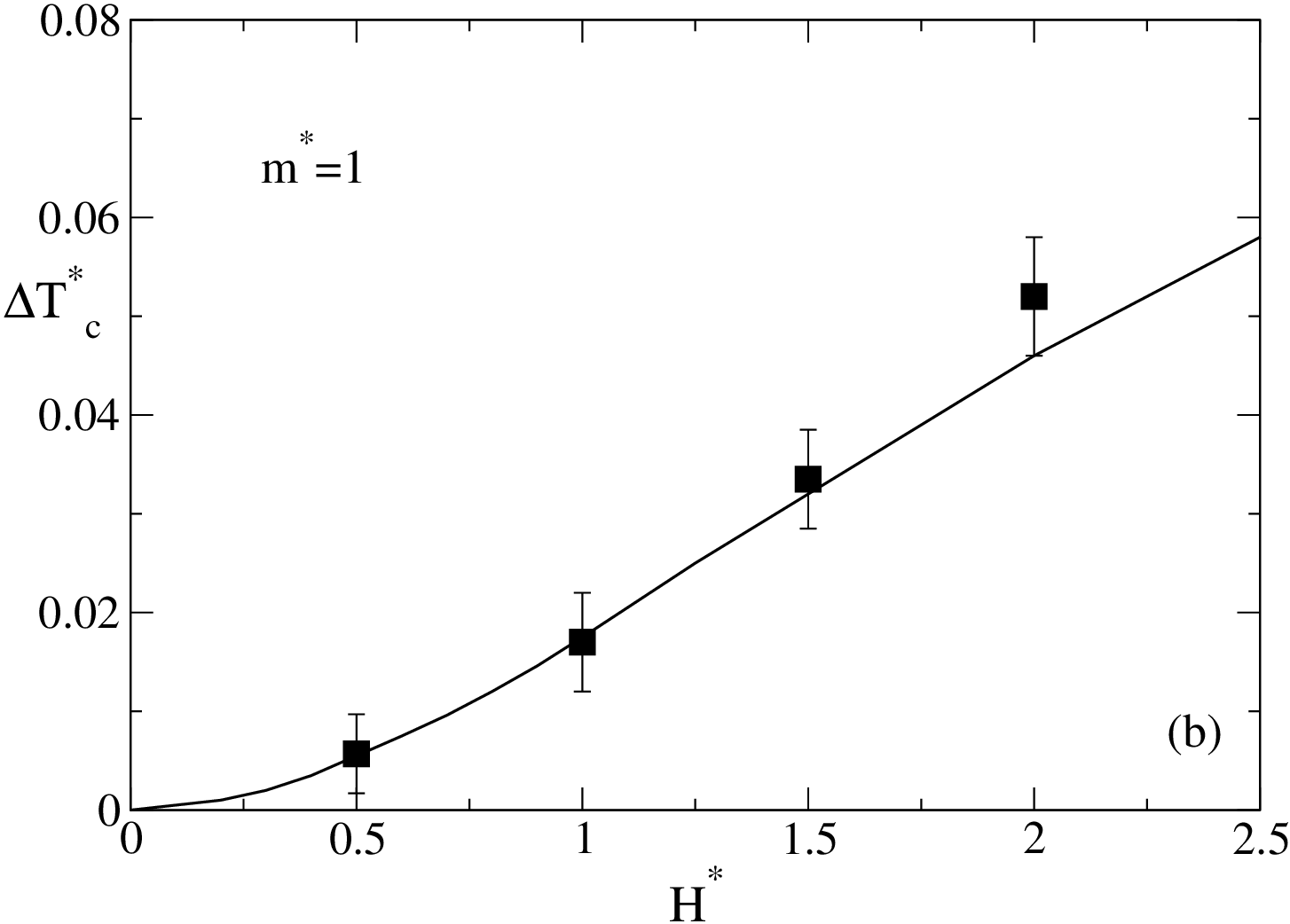}
\end{minipage}
\begin{minipage}{37pc}
\caption{\label{label}(a) Evolution of the phase diagram as function of the external 
magnetic field for a DY fluid in a needle-shaped container for $m^*=1$. 
(b) Shift of the critical 
temperature ($\Delta{T^*_c}=T^*_c(H)-T^*_c(0)$) of the liquid-vapor critical point 
as a function of the 
applied field for $m^*=1$. The symbols in both figures are the results of 
Monte Carlo simulations whereas the lines are obtained from the present DFT.} 
\end{minipage}
\end{figure}
for a needle-shaped sample are displayed in Fig. 1(a) together with the result of the 
Langevin theory for $m^*=1$. 
Figure 1(a) shows that the interparticle interaction enhances the magnetization  
relative to the corresponding values of the Langevin theory. The theoretical data 
of the present theory are consistent with our new simulation data. For the reduced 
density $\rho^*=0.25$ we find both DFT and MC results to be in excellent, quantitative 
agreement for all field strengths. For $\rho^*=0.6$ the level of quantitative 
agreement reduces for larger field strengths. 

The density dependence of the magnetic susceptibility obtained from the numerical 
solution of Eqs. (\ref{magn2}) and (\ref{sus}) in comparison with earlier 
simulation data of Szalai {\it et al} \cite{s1} is displayed in Fig. 1(b).
For the reduced dipole moment $m^*=1$ the agreement between the theoretical and 
simulation results is quite good, especially at lower densities. For $m^*=\sqrt{2}$ 
the level of quantitative agreement reduces for larger reduced densities.

The liquid-vapor phase diagrams at a given 
temperature $T$ follow from requiring the equality of the chemical potential and of 
the pressure for the densities of coexisting phases. 
The phase diagrams in the density-temperature plane are
presented  in Fig. 2(a) for various fixed values of the magnetic field and for the reduced 
dipole moment $m^*=1$. Figure 2(a) shows that for $H^*=0$ and $m^*=1$ there is no 
spontaneous magnetization within the 
framework of this theory, i.e., there is no ferromagnetic liquid phase. 
For $H^*\neq{0}$ there is a first-order phase transition 
between a low density gas and a high density liquid phase and both phases 
exhibit a nonzero magnetization. The agreement between the NpT+TP simulation 
and the DFT data is reasonable. 
For $m^*=1$ Fig. 2(b) represents the magnetic field 
dependence of the critical temperature corresponding to the liquid-gas phase transitions, 
obtained from the present DFT and liquid-gas equilibrium NpT+TP simulation data. 
For small $H^*$ both the theoretical and the simulation shifts 
increase quadratically as a function of the field strength. Similar results have been 
found for Stockmayer fluids in Refs. \cite{g2,b1}.  

\section{Summary}
The following main results have been obtained.

(1) Based on a second-order Taylor series expansion of the free energy functional of
the dipolar Yukawa fluids, within the mean spherical approximation we have derived an 
analytical expression for the magnetization of ferrofluids. There is good agreement 
between the results of the density functional theory and the Monte Carlo simulation data 
for reduced dipole moments $m^*\lesssim 1.4$. The proposed implicit equation 
for the magnetization (Eq. (\ref{magn2})) extends the applicability of MSA for arbitrary 
strengths of the magnetic fields.  

(2) We have derived analytical equations for the external field dependence of
the pressure (Eq. (\ref{pres})) and of the chemical potential
(Eq. (\ref{chem})) for dipolar Yukawa fluids which allows one to determine the 
magnetic field dependence  of the liquid-vapor phase equilibria. 
\ack{} 
I. Szalai would like to thank the Hungarian Scientific Research Fund 
(Grant No. OTKA K61314) for financial support.




\section*{References}
\begin{thebibliography}{9}
\bibitem{bu} Buyevich Y A and Ivanov A O 1992 {\it Physica} A {\bf 190} 276
\bibitem{hu} Huke B and L\"ucke 2004 {\it Rep. Prog. Phys.} {\bf 67} 1731
\bibitem{ba} Russier V and Douzi M 1994 {\it J. Colloid Interface Sci.} 
    {\bf 162} 356 
\bibitem{g1} Groh B and Dietrich S 1994 {\it Phys. Rev.} E {\bf 50} 3814
\bibitem{sa1} Szalai I and Dietrich S 2005 {\it Mol. Phys.} {\bf 103} 2873
\bibitem{kl1} Range G M and Klapp S H L 2004 {\it Phys. Rev.} E {\bf 70} 031201
\bibitem{h1} Henderson D, Boda D, Chan K-Y and Szalai I 1999 {\it
    J. Chem. Phys.} {\bf 110} 7348
\bibitem{s1} Szalai I, Henderson D, Boda D and Chan K-Y 1999 {\it
    J. Chem. Phys.} {\bf 111} 337
\bibitem{w1} Wertheim M S 1971 {\it J. Chem. Phys.} {\bf 55} 4291
\bibitem{l1} Lebedev A V 1989 {\it Magnetohydrodynamics} {\bf 25} 520
\bibitem{m1} Morozov K I and Lebedev A V 1990 {\it J. Magn. Magn. Mater.} {\bf 85} 51
\bibitem{l2} Lovett R, Mou C Y and Buff F P 1976 {\it J. Chem. Phys.} {\bf 65} 570
\bibitem{ha1} Hausleitner C Hafner J 1988 {\it J. Phys.} F: {\it Met. Phys.} {\bf 18} 1013
\bibitem{pe1} Percus J K and Yevick G J 1956 {\it Phys. Rev.} {\bf 110} 1
\bibitem{t1} Tang Y 2003 {\it J. Chem. Phys.} {\bf 118} 4141 
\bibitem{b2} Boda D, Liszi J and Szalai I 1995 {\it Chem. Phys. Lett.} {\bf 235} 140
\bibitem{g2} Groh B and Dietrich S 1996 {\it Phys. Rev.} E {\bf 53} 2509
\bibitem{b1} Boda D, Winkelmann J, Liszi J and Szalai I 1996 {\it Mol. Phys.} 
    {\bf 87} 601 
\end{thebibliography}
\end{document}